\documentclass[runningheads]{llncs}
\usepackage{graphicx}
\usepackage{amsmath} % enabling maths symbols and equations
\usepackage{amsfonts} % enabling extra maths fonts
\usepackage{amssymb} % enabling extra maths symbols
\usepackage{enumitem}
\usepackage{latexsym}
\usepackage{newtxmath}

\begin{document}
\title{Cyber-Security Investment in the Context of Disruptive Technologies: Extension of the Gordon-Loeb Model}
\titlerunning{Cyber-Security Investment in the Context of Disruptive Technologies}
\author{Dimitri Percia David\inst{1}\orcidID{0000-0002-9393-1490} \and
Alain Mermoud\inst{1}\orcidID{0000-0001-6471-772X} \and
Sébastien Gilliard\inst{1}\orcidID{n.a.}}
\authorrunning{D. Percia David et al.}
\institute{Military Academy at ETH Zurich, Department of Defense Economics}

\maketitle              % typeset the header of the contribution
\begin{abstract}
Cyber-security breaches inflict significant costs on organizations. Hence, the development of an information-systems defense capability through cyber-security investment is a prerequisite. The question of how to determine the optimal amount to invest in cyber-security has been widely investigated in the literature. In this respect, the Gordon-Loeb model and its extensions received wide-scale acceptance. However, such models predominantly rely on restrictive assumptions that are not adapted for analyzing dynamic aspects of cyber-security investment. Yet, understanding such dynamic aspects is a key feature for studying cyber-security investment in the context of a fast-paced and continuously evolving technological landscape. We propose an extension of the Gordon-Loeb model by considering multi-period and relaxing the assumption of a continuous security-breach probability function. Such theoretical adaptations enable to capture dynamic aspects of cyber-security investment such as the advent of a disruptive technology and its investment consequences. Such a proposed extension of the Gordon-Loeb model gives room for a hypothetical decrease of the optimal level of cyber-security investment, due to a potential technological shift. While we believe our framework should be generalizable across the cyber-security milieu, we illustrate our approach in the context of critical-infrastructure protection, where security-cost reductions related to risk events are of paramount importance as potential losses reach unaffordable proportions. Moreover, despite the fact that some technologies are considered as disruptive and thus promising for critical-infrastructure protection, their effects on cyber-security investment have been discussed little.

\end{abstract}
\newpage
\section{Introduction}
Cyber-security\index{cyber-security} breaches inflict significant costs on organizations, businesses, and individuals \cite{anderson_measuring_2013,campbell_economic_2003,gordon_2005a}. As a result, operators of information systems (IS) must develop an IS-defense capability\index{information-systems (IS) defense capability}. In order to build a capability, investing in material resources is a prerequisite \cite{teece_capability_2019,barney_firm_1991,penrose_theory_2009}. In the context of cyber-security, such material resources are composed by processes, products and/or services that are related to the security management of IS in general and, among others, to cyber-threats monitoring in particular \cite{amini_2015,schatz_economic_2017,singh_2013,anderson_2010security} -- e.g., intrusion-detection systems (IDS). 

The cyber-security investment issue has received significant academic attention (for an extensive literature review, see \cite{schatz_economic_2017}), and such a literature is often directly related to practitioners' needs. For instance, as organizations face budget constraints, an important challenge for these same organizations is to maximize the efficiency of any monetary investment in cyber-security processes, products and/or services \cite{bohme_economics_2013,gordon_economics_2002,schatz_economic_2017}. Prior IS research has produced many quantitative models that propose to optimize such investment, as well as recommendations to invest in particular technologies or systems (e.g., \cite{bohme_economics_2013,herath2008,jerman2012,li2007,shirtz2011}). These formal approaches are complemented by less formal practitioner-oriented discussions \cite{azorin2019,smeraldi2014,weinberger2011}. Among all the available models and approaches, the Gordon-Loeb (GL) model has received wide-scale acceptance \cite{zhou_2018}. The GL model is based on microeconomics -- more precisely on the fundamental economic principle of cost-benefit analysis \cite{gordon_externalities_2015,gordon_economics_2002} --, establishing a general setup for determining the optimal level of cyber-security investment\index{investment}. By specifying a security-breach probability function (SBPF), the GL model attempts to determine the above-mentioned optimal investment amount for cyber-security processes, products and/or services (e.g., \cite{baryshnikov_it_2012,gordon_externalities_2015,lelarge_coordination_2012,gordon_economics_2002,willemson_extending_2010}).

However, the GL model framework evades dynamic issues such as perverse economic incentives\footnote{ For example, externalities arising when the decisions of one party affect those of others \cite{anderson_why_2001}.} and the advent of a \textit{disruptive}\footnote{ In economic terms, the notion of \textit{disruptive technology} \cite{christensen_what_2015} refers to a radically innovative technology that significantly disrupts existing economic structures, markets and value networks, thus displacing established leading products, processes and/or services \cite{christensen_what_2015}. Therefore, a \textit{disruptive technology} comes from innovation. However, not all innovations are disruptive, even though they can be revolutionary. For instance, the advent of personal computers (PC) in the IT landscape can be considered as a disruption. The multi-purpose functionalities of PCs, as well as their relatively small size, their extended capabilities, and their low price facilitated their spread and individual use. PCs displaced large and costly minicomputers and mainframes, significantly affecting the lives of individuals and the management of organizations.} cyber-security\index{cyber-security} technology\index{information technologies (IT)} \cite{gordon_economics_2002}. Yet, in the case of the latter, accounting for such dynamic issues could significantly enrich the initial model by giving supplementary time-related insights concerning the potential consequences that disruptive technologies might trigger for cyber-security investment\index{investment}. If conventional cyber-security processes, products and/or services such as IDS are essentially assimilated to the \textit{targeted event-based detection} approach \cite{amini_2015,singh_2013,anderson_2010security}, the swift evolution of both the technological landscape and cyber-threats calls for a complementary approach based on \textit{behavior-anomaly detection} \cite{amini_2015,singh_2013,wang_2014a,casas_2017,cui_2016,sait_2015,terzi_2017}, and its subsequent (potentially) disruptive technologies. 

By extending the original GL model\index{Gordon-Loeb (GL) model} to a multi-period setup, and by relaxing the assumption of a continuously twice-differentiable SBPF\index{security-breach probability function (SBPF)}, we create room for additional insights on cyber-security investment\index{investment} by delivering a theoretical ground that enables to investigate the effect of a disruptive technology\index{disruptive technologies} on such investment. In this article, we illustrate the concept of disruptive technologies\index{disruptive technologies} through the case of big-data analytics\footnote{ In this article, the term \textit{big data} refers to data whose complexity impedes it from being processed (mined, stored, queried and analyzed) through conventional data-processing technologies\index{information technologies (IT)} \cite{laney_3d_2001}. The complexity of big data is defined by three attributes: $(1)$ the volume (terabytes, petabytes, or even exabytes ($10^{18}$ bytes); $(2)$ the velocity (referring to the fast-paced data generation); and $(3)$ the variety (referring to the combination of structured and unstructured data) \cite{laney_3d_2001}. The field of BDA is related to the extraction of value from big data -- i.e., insights that are non-trivial, previously unknown, implicit and potentially useful. BDA extracts patterns of actions, occurrences, and behaviors from big data by fitting statistical models to these patterns through different data-mining techniques (e.g., predictive analytics, cluster analysis, association-rule mining, and prescriptive analytics) \cite{sathi_big_2012,cardenas_big_2013}.} (BDA) technologies and their related techniques. To the best of our knowledge, our proposed extension of the GL model is the first approach responding to the need of capturing the aforementioned time-related insights that a disruptive technology might bring on cyber-security investment.

The remainder of this article is structured as follows. In Section 2, we contextualize this research by emphasizing the evolution of cyber-threats monitoring approaches and their potential disruptive technologies. In Section 3, we present an extension of the GL model\index{Gordon-Loeb (GL) model} and we provide related propositions in order to create room for capturing the consequences of disruptive technologies\index{disruptive technologies} on cyber-security investment\index{investment}. In Section 4, we suggest that our framework can be applied by critical-infrastructure providers (CIP)\index{critical-infrastructure providers (CIP)} in order to help them optimize their investment\index{investment} in IS defense\index{information-systems (IS) defense capability} technologies\index{information technologies (IT)}. In the last section, we discuss limitations and future work.

\section{Cyber-Threat Monitoring Approaches}
\label{Isec2}

Cyber-threat monitoring approaches can be divided in two broad classes: (1) \textit{targeted event-based detection} and (2) \textit{behavior-anomaly detection} \cite{amini_2015,singh_2013}. Both approaches aim to detect suspicious events (i.e., related to security concerns such as data breaches) through IDS \cite{singh_2013}. However, technologies related to (2) are rapidly evolving and could disrupt the cyber-security market and the investment in these technologies.

    \subsection{The \textit{Targeted Event-Based Detection} Approach}
    \label{Isubsec2.1}
    For the past three decades, scholars, practitioners, and organizations\index{organization} have been developing numerous conventional technical means to increase cyber-security\index{cyber-security} by using signature-detection measures and encryption techniques \cite{anderson_2010security,anderson_economics_2006}. However, the success of such conventional approaches has been limited \cite{chen_business_2012,anderson_2010security,anderson_economics_2006,bohme_economics_2013}. 
    
    The \textit{targeted event-based detection} approach essentially relies on signature-detection measures in which an identifier is attributed to a known threat\footnote{ Common cyber-risks include malware (spyware, ransomware, viruses, worms, etc.), phishing, man-in-the-middle attacks (MitM), (distributed) denial-of-service attacks ((D)DoS), malicious SQL injections, cross-site scripting (XSS), credential reuse, and brute-force attacks.} (i.e., a threat that had been witnessed in the past), so that this threat can be subsequently identified \cite{anderson_2010security,singh_2013,amini_2015}.  Examples of \textit{targeted event-based detection} include network-IDS, access-control mechanisms, firewalls and pattern-based antivirus engines \cite{amini_2015,singh_2013,anderson_2010security}. For instance, a signature-detection technology such as an anti-virus scanner might detect a unique pre-established pattern of code that is contained in a file. If that specific pattern (i.e., signature) is discovered, the file will be flagged in order to warn the end-user that their file is infected. \cite{anderson_2010security}. Yet, such an approach is becoming more and more ineffective, as the swift evolution of cyber-threats is becoming more sophisticated \cite{anderson_2010security,amini_2015,wang_2014a,casas_2017}. Over the last decade, cyber-crimes have rapidly increased because hackers have developed new procedures to circumvent IS security to gain unauthorized and illegal access to the system. For instance, as malware spread from one system to the next, hackers employ \textit{polymorphic code}\footnote{ A \textit{polymorphic code} is a code that employs a polymorphic engine in order to mutate while keeping the original algorithm intact. In other terms, the code changes itself each time it runs, but its semantics -- the function of the code -- will not change \cite{forest_2002}.} techniques in order to (automatically) rapidly modify the pattern \cite{forest_2002}, thus evading and circumventing existing detection mechanisms.\footnote{ Additional techniques such as, sandbox resistance, fast fluxing, adversarial reverse engineering, social-engineering attacks, spoofing, and advanced persistent threats (APT) are continuously evolving and spreading \cite{amini_2015}.} Consequently, such a \textit{targeted event-based detection} would be ineffective because of the lag that exists between the development of signatures and the rapid expansion of cyber-threats. Moreover, \textit{zero-day}\footnote{ A \textit{zero-day} vulnerability is a computer-software vulnerability that is either unknown to or unaddressed by operators who should be interested in mitigating the vulnerability. \textit{Zero-day} vulnerabilities enable hackers to exploit it in order to adversely affect computer programs, networks, and data \cite{holm_2014}.} vulnerabilities cannot be caught with such an approach.

    Therefore, an important limiting factor of the \textit{targeted event-based detection} approach is that it is intrinsically reactive in nature \cite{anderson_2010security,amini_2015}. Attributing a signature to a cyber-threat is always preceded by a cyber-incident, which implies that signatures are unable to identify unknown and/or emerging threats. As a consequence, signature-detection measures can be rendered almost ineffective by hackers \cite{wang_2014a,casas_2017}. Such a scenario is even amplified in the era of extended digitization and big data \cite{singh_2013}. The reasons are that (1) the swift development of computer networks (e.g., the extensive use of cloud and mobile computing) in the past decades generated new channels that expose data to cyber-attack, thus increasing the pool of systems to be attacked hence amplifying numerous security issues related to intrusions on computer and network systems \cite{singh_2013}; (2), up to multiple exabytes of information are being transferred daily, usually impeding the process of the entire set of security information, e.g., network logs, access records, etc.; (3) the velocity of data generation makes any type of data processing difficult through conventional computer hardware and software architectures; (4) the complexity of data also impedes security information from being processed by conventional computers, e.g., data come from diverse sources, it is stored in different formats and on different IT. Hence, the damage done by cyber-threats could be identified only after an attack, giving hackers more efficient possibilities to access networks, to hide their presence and to inflict damage \cite{chen_business_2012}. 

    \subsection{The \textit{Behavior-Anomaly Detection} Approach}
    \label{Isubsec2.2}
    In order to address the drawbacks of the \textit{targeted event-based detection} approach, research \& development has been focused on a \textit{behavior-anomaly detection} approach \cite{chen_business_2012,singh_2013,sait_2015,cui_2016,casas_2017,terzi_2017,wang_2014a}. Such an approach aims to detect suspicious/unusual events through the extraction of patterns, by using behavior anomalies and/or deviations from behaviors (actions, occurrences) of entities and/or users of a network, rather than patterns of code as in the case of the \textit{targeted event-based detection} \cite{chen_business_2012,singh_2013,sait_2015,cui_2016,casas_2017,terzi_2017,wang_2014a}. The \textit{targeted event-based detection} approach is predominantly based on \textit{security analytics}\footnote{ The field of \textit{security analytics} aims to detect suspicious events by extracting patterns related to behavioral anomalies and/or deviations from behaviors (actions, occurrences) of entities and/or users of a network. In other words, \textit{security analytics} methods aim to distinguish patterns generated by legitimate users from patterns generated by suspicious and/or malicious users \cite{chen_business_2012}.} and aims to provide dynamic detection of cyber-threats through techniques derived from BDA by fitting statistical models on these patterns through different techniques (e.g., predictive analytics, cluster analysis, association-rule mining, and prescriptive analytics) \cite{sathi_big_2012,cardenas_big_2013} used for network forensics, traffic clustering and alert correlation \cite{amini_2015}. Some examples have been developed by \cite{terzi_2017,cui_2016,sait_2015}. 
    
    Complementing the \textit{targeted event-based detection} approach by using signature-detection measures, such a \textit{behavior-anomaly detection} approach is envisioned to foster cyber-threats monitoring by increasing its productivity \cite{chen_business_2012}. Examples of \textit{security analytics} applications are the development of \textit{security information and event management} (SIEM). SIEM products and services combine security-information management (SIM) and security-event management (SEM) \cite{bhatt_2014}. SIEM provide real-time analysis of information-security alerts generated by applications and network hardware. SIEM are sold as managed services, as software, or as appliances \cite{bhatt_2014}. These products and services are also used to log security data and to generate reports for compliance purposes. SIEM products and services are widely used within \textit{information-security operation centers} (ISOC) -- also called \textit{fusion centers} -- of organizations. The open-threat exchange platform called \textit{AlienVault OTX} is an example of SIEM products and services.\footnote{ \url{https://www.alienvault.com/}}

    Cyber-security technologies related to the \textit{behavior-anomaly detection} approach -- and more precisely related to \textit{security analytics} -- are susceptible to disrupting the cyber-security market \cite{wang_2014a,terzi_2017,cui_2016,sait_2015}. However, the dynamic (i.e., time related) consequences of such disruptive technologies on cyber-security investment cannot be analyzed through the existing GL model and its extensions. They rely on restrictive assumptions that are not adapted for analyzing dynamic aspects of cyber-security investment \cite{gordon_economics_2002}. Consequently, a supplementary extension of the GL model is necessary.

\section{Extending the Gordon-Loeb Model}
\label{Isec3}

\begin{figure}[h]
        \includegraphics[scale=0.75]{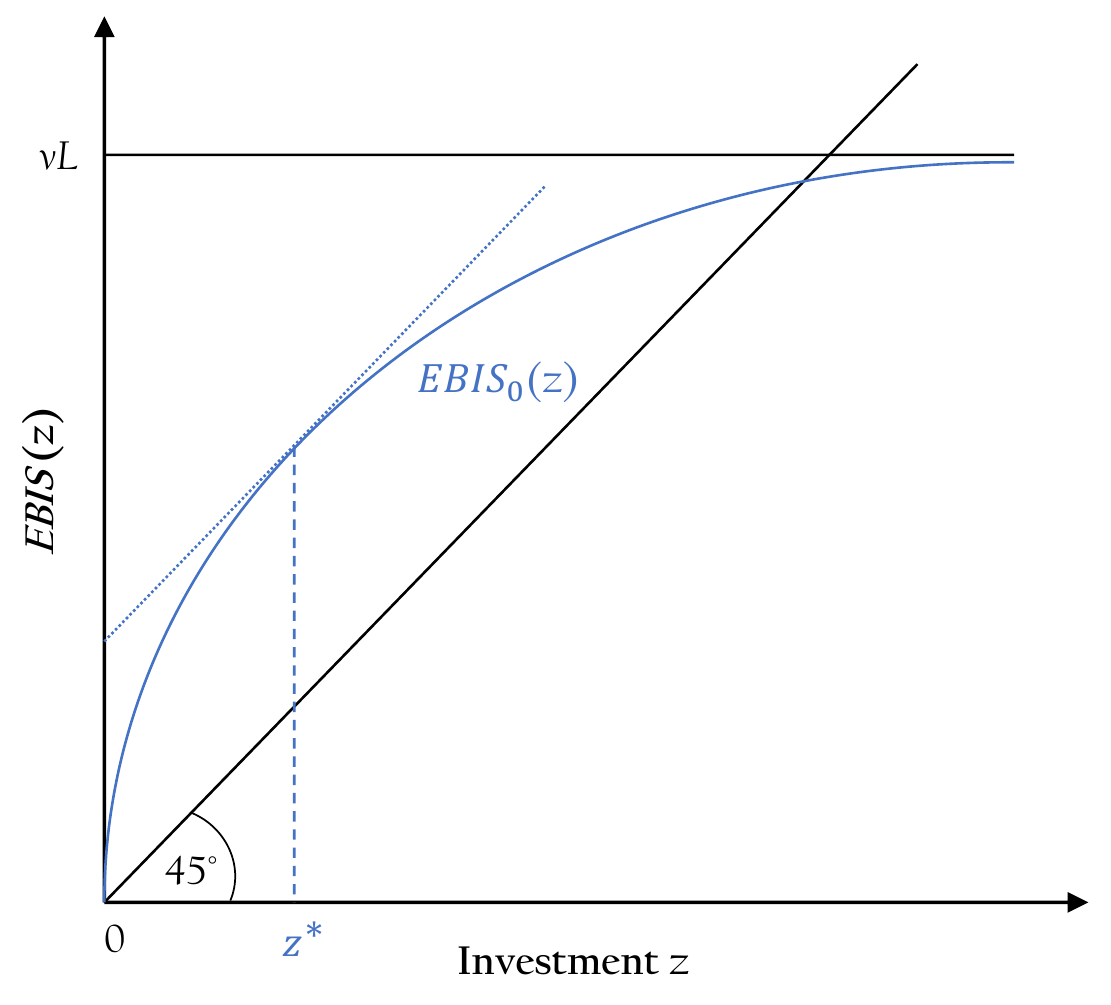}
        \caption{\textbf{Level of Investment in Cyber-Security}\\ Notes to Figure \ref{figI.1}: The original function of the \textit{expected benefits in cyber-security} ($EBIS_{0}(z)$) has a do- main $Z$ that yields a distribution between zero and the monetary value of the expected loss ($vL$) in the absence of any cyber-security investment. The monetary value (i.e., costs) of the investment ($z$), corresponds to the 45° line. The optimal level of cyber-security investment ($z^{*}$) is obtained when the difference between benefits and costs is maximized (tangent to $EBIS_{0}(z)$ -- where the marginal benefits are equivalent to the marginal cost of one -- yielding a slope of 45°, in blue). \textit{N.B.}: the optimal level of cyber-security investment ($z^{*}$) is smaller than the expected loss ($vL$) in the absence of any investment.} 
        \label{figI.1}
\end{figure}

\begin{figure}[h]
        \includegraphics[scale=0.75]{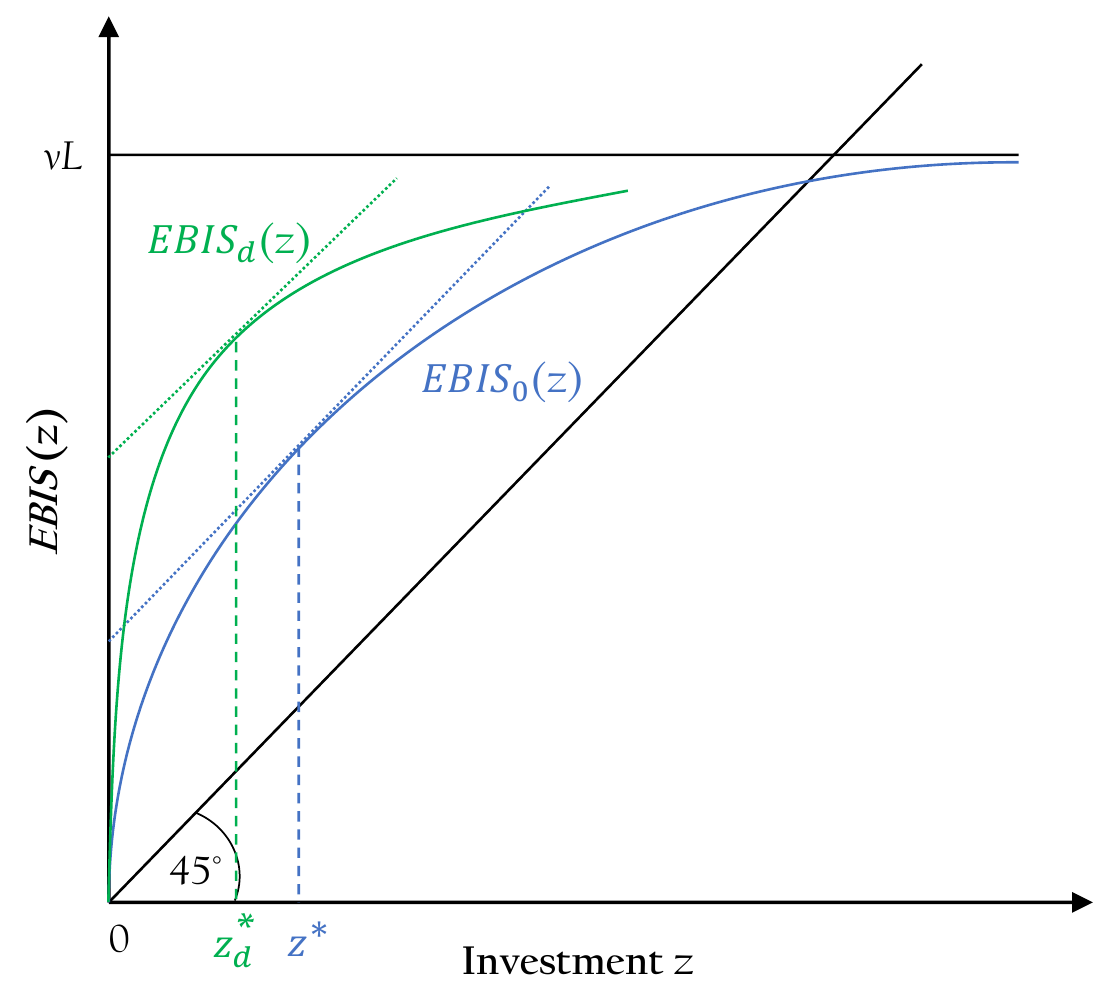}
        \caption{\textbf{Left Shift of the Level of Investment in Cyber-Security}\\ Notes to Figure \ref{figI.2}: The new function of the \textit{expected benefits in cyber-security} ($EBIS_{d}(z)$) has also a do- main $Z$ that yields a distribution between zero and the monetary value of the expected loss ($vL$) in the absence of any cyber-security investment. The monetary value (i.e., costs) of the investment ($z$), also cor- responds to the 45° line. Similarly to Figure \ref{figI.1}, the optimal level of cyber-security investment ($z^{*}_{d}$) is obtained when the difference between benefits and costs is maximized (tangent to $EBIS_{d}(z)$, yielding a slope of 45°, in green). Yet, due to the left shift of the $EBIS$ function (from $EBIS_{0}(z)$ to $EBIS_{d}(z)$), the optimal level of cyber-security investment also shifts from $z^{*}$ to $z^{*}_{d}$.} 
        \label{figI.2}
\end{figure}

\begin{figure}[h]
        \includegraphics[scale=0.75]{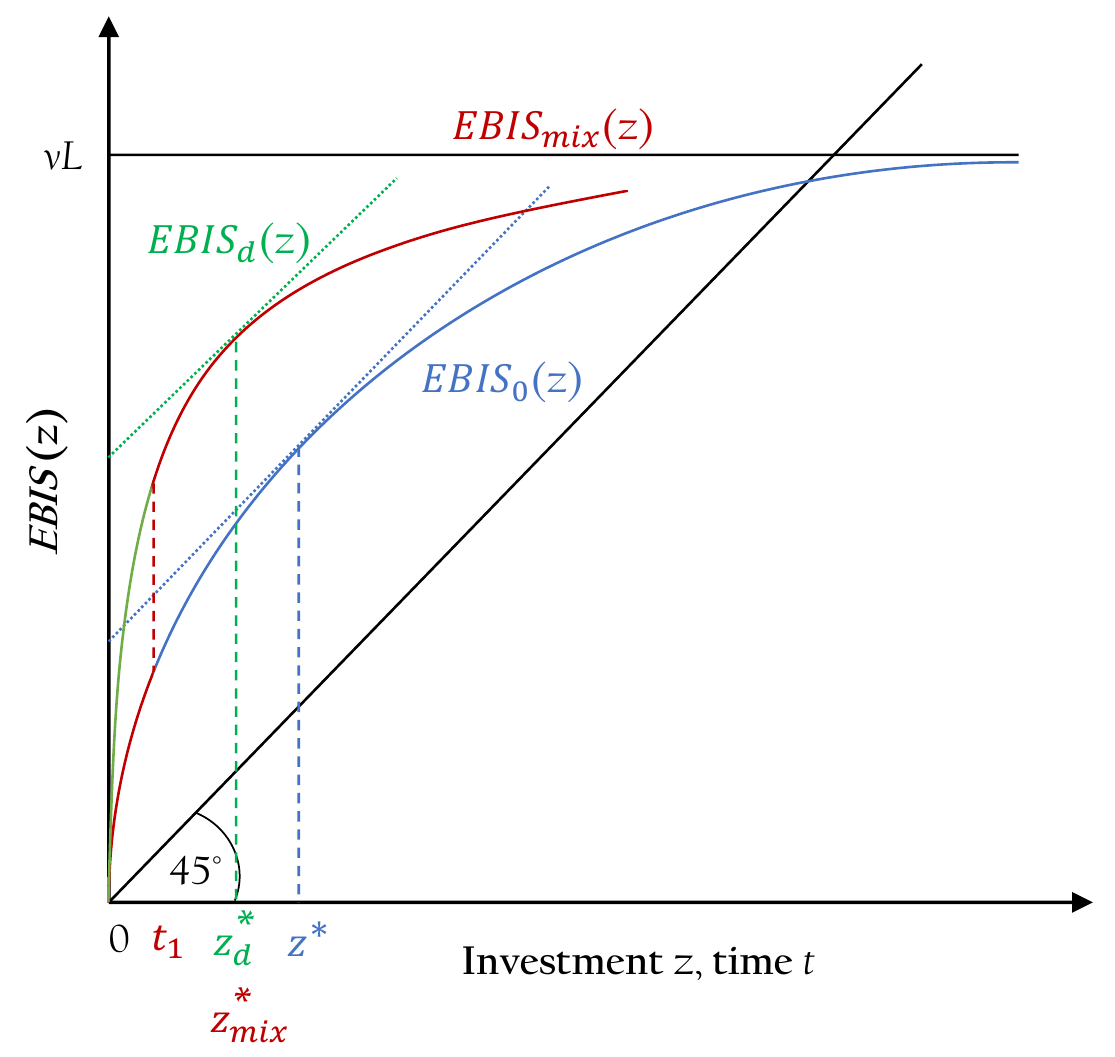}
        \caption{\textbf{Left Shift of the Level of Investment in Cyber-Security Over Time}\\ Notes to Figure \ref{figI.3}: In this graph, the horizontal axis represents both investment ($z$) and time ($t$). The red line represents the function of the \textit{expected benefits in cyber-security} over time ($EBIS_{mix}(z)$). The red dashed line represents the discontinuity in the domain $Z$ of the $EBIS_{mix}(z)$ function -- due to the implementation of a disruptive technology at time $t_{1}$. Without the implementation of a disruptive technology, the function of the \textit{expected benefits in cyber-security} remains $EBIS_{0}(z)$ (in blue). With the implementation of a disruptive technology from $t=0$, the function of the \textit{expected benefits in cyber-security} is $EBIS_{d}(z)$ (in green). The optimal level of cyber-security investment is $z^{*}_{mix} = z^{*}_{d}$.}  

        \label{figI.3}
    \end{figure}

By focusing on costs and benefits associated with cyber-security\index{cyber-security}, the GL model\index{Gordon-Loeb (GL) model} states that each organization’s objective\index{organization} is to maximize its \textit{expected net-benefits in cyber-security} function ($ENBIS$) \cite{gordon_economics_2002}. This corresponds to minimizing the total expected cost, equivalent to the addition of the expected loss\footnote{ Wherein $v$ is is the organization’s inherent \textit{vulnerability} -- defined as the probability that a threat, once realized (i.e., an attack), would be successful -- to cyber-security breaches, ($\textit{P}$); and $L$ is the potential loss associated with the security breach, (\$). The model description and its assumptions were explained in detail by \cite{gordon_externalities_2015,gordon_economics_2002}.} ($vL$) due to cyber-security breaches and the expenses ($z$) in cyber-security processes, products and/or services implemented and/or undertaken in order to counter such breaches \cite{gordon_externalities_2015,gordon_economics_2002}. Figure \ref{figI.1} on page~\pageref{figI.1} illustrates the maximization of the expected benefits coming from cyber-security expenditures ($EBIS$).\footnote{ In order to simplify the illustration, figures \ref{figI.1}, \ref{figI.2} and \ref{figI.3} on pages~\pageref{figI.1} and \pageref{figI.3} depicts the $EBIS$ function instead of the $ENBIS$ function. The $ENBIS$ function is obtained by subtracting the investment ($z$) to the $EBIS$ function.} 

In Section \ref{Isec2}, we emphasized that the advent of the \textit{behavior-anomaly detection} approach -- triggered by \textit{security analytics} and its use of BDA -- is bringing to the fore new cyber-security technologies that have the potential to disrupt the cyber-security market \cite{wang_2014a,terzi_2017,cui_2016,sait_2015,sathi_big_2012} by hypothetically bringing superior returns on investment\index{investment} \textit{vis-à-vis} conventional measures related to the \textit{targeted event-based detection} approach \cite{sathi_big_2012}. As a consequence, the $EBIS$ function might shift to the left. Figure \ref{figI.2} on page~\pageref{figI.2} illustrates the potential maximization of the $EBIS$ function in the context of BDA. Accordingly:
 
	\begin{quote}
		\textit{\textbf{Proposition 1:} If BDA is employed in order to provide cyber-threats monitoring\index{cyber-security}, the $ENBIS$ function will shift to the left due to a greater productivity\index{productivity}\index{productivity} of BDA compared to conventional technologies.}
	\end{quote}

\noindent
Consequently, the optimal level of cyber-security investment will decrease from $z^{*}$ to $z^{*}_{d}$ (c.f.: Figure \ref{figI.2} on page~\pageref{figI.2}). For the same level of protection, investment in cyber-security will decrease -- \textit{ceteris paribus}. Accordingly:
 
	\begin{quote}
		\textit{\textbf{Proposition 2:} If BDA is implemented in order to provide cyber-threats monitoring\index{cyber-security}, the $ENBIS$ function will witness a discontinuity in its domain $Z$ due to a greater productivity\index{productivity}\index{productivity} of BDA compared to conventional technologies.}
	\end{quote}

\newpage

\noindent
Figure \ref{figI.3} on page~\pageref{figI.3} illustrates such a discontinuity in the $ENBIS$ function described in Proposition 2. Yet, as the implementation of a disruptive technology implicitly triggers a productivity\index{productivity} differential over time, we propose to extend the GL model\index{Gordon-Loeb (GL) model} in two distinct but related ways.

\vspace{0.3cm}

    \subsection{A Temporal Setup}
    First, in order to capture the advent of a disruptive technology\index{disruptive technologies} and its dynamic consequences on cyber-security investment\index{investment}, a temporal setup has to be implemented. As the GL model\index{Gordon-Loeb (GL) model} was developed for a single-period, it excludes the fundamental temporal dimension for analyzing the technological-shift\index{information technologies (IT)} dynamics induced by efficiency\index{efficiency} improvements. Hence, the extension of the original single-period model to a multi-period\footnote{ Even though there is still an open debate in the field of mathematics about whether time should be considered as continuous or discrete \cite{serfozo_1979}, in our proposed extension of the GL model, we consider time as discrete. Such a choice is determined by the fact that any further empirical research that would test our extension will be essentially conditioned by measuring time as a discrete variable. Consequently, the following extension is based on an antidifference instead of an antiderivative.} setup might bring significant insight in understanding the dynamic aspects of cyber-security investment\index{investment} that were originally evaded. 
    
    Specifically, we adapt the GL model\index{Gordon-Loeb (GL) model} from \cite{gordon_economics_2002}, such as the maximization of the $ENBIS$ function is determined by an antidifference operator of this same function, which has the domain $Z \subseteq \mathbb{R}_{\geq 0}$ (representing the set of investment possibilities, $z_{i}$) at the end of the specified time horizon $T \subseteq \mathbb{N}$, wherein each period $i \in [1, n]$. This yields:
    
	        \vspace{0.5cm}
 		    
	    \begin{equation} \label{eq:1}
		    \max_{z_{i} \in Z} \Delta^{-1} ENBIS(z_{i}) = \max_{z_{i} \in Z} {\left\{\sum_{i=1}^{n} [v_{i}-S^{I}_{i}(z_{i,}v_{i})]L_{i}-z_{i}\right\}}
	    \end{equation}
	    
 		    \vspace{0.2cm}

	    \begin{quote}
		    wherein for each period $i \in [1,n]$:
   			    \begin{itemize}
			    	\item[--] $v_{i}$ is the organization’s\index{organization} \textit{vulnerability}\footnote{ In the original GL model \cite{gordon_economics_2002}, the organization's inherent vulnerability is defined as the probability that a threat, once realized (i.e., an attack), would be successful.} to cyber-security breaches, (probability $\mathbb{P}$); 
				    \item[--] $S^{I}_{i}$  is the organization’s \textit{security-breach probability function} (SBPF)\index{security-breach probability function (SBPF)}, defined as the probability that a cyber-security-breach occurs\index{cyber-security}, (probability $\mathbb{P}$); 
				    \item[--] $z_{i}$ is the organization’s investment\index{investment} in cyber-security\index{cyber-security}, (monetary value \$);
				    \item[--] $L_{i}$ is the potential loss associated with the security breach, (monetary value \$).
   			    \end{itemize}
	    \end{quote}

    \subsection{A Discontinuity in the Security-Breach Probability Function}
    Second, the presumed technological\index{information technologies (IT)} shift induced by the superior productivity\index{productivity} of a disruptive technology\index{disruptive technologies} challenges the assumption of a continuously twice-differentiable security-breach probability function\index{security-breach probability function (SBPF)}. The original model defines continuously decreasing but positive returns to scale of cyber-security\index{cyber-security} investment\index{investment}. Yet, this continuously twice-differentiable setup leaves no room for a discrete emergence of a technological\index{information technologies (IT)} shift brought by a disruptive and more efficient technology.\footnote{ \cite{gordon_economics_2002} explicitly acknowledge that they ‘abstract from reality and assume that postulated functions are sufficiently smooth and well behaved', thus creating favorable conditions for applying basic differential calculus, hence simplifying the optimization problem of the security-investment\index{investment} phenomenon. Although a smooth approximation of the security-investment\index{investment} phenomenon done by \cite{gordon_economics_2002} is a reasonable first approach, in order to deliver insight concerning the problem of determining an optimal level of cyber-security\index{cyber-security} investment\index{investment}, such an approach lacks realism. As explicitly mentioned by \cite{gordon_economics_2002}: [...] ‘\textit{in reality, discrete investment\index{investment} in new security technologies\index{information technologies (IT)} are often necessary to get incremental result. Such a discrete investment\index{investment} results in discontinuities.}'} In such a theoretical framework, the elasticity of the protection of cyber-security\index{cyber-security} processes, products and/or services evades radical technological\index{information technologies (IT)} progress. However, technological progress induced by the implementation of a disruptive technology -- such as BDA -- could considerably reduce cyber-security investment\index{investment} by bringing suggestively greater returns on investment\index{investment}.\footnote{ In BDA, an extremely large, fast paced and complex amount of information can be processed in significantly shortened time frames and at almost zero marginal cost per additional unit of information -- once the fixed development and implementation costs of systems and algorithms for investigating threat patterns are invested \cite{sowa_conceptual_1983}. Furthermore, the real-time analytics provided by big-data algorithms are likely to neutralize any attacker’s information advantage, such that the probability of a cyber-breach should be reduced. For example, an attacker can exploit \textit{zero-day} vulnerabilities by knowing where to attack, whereas the defender does not know, hence has to protect all potential entry spots. As real-time analytics reveals both the time and the position of the attack as it happens, the defender can react precisely on the attacked spot, thus save any unnecessary investment\index{investment} in the protection of spots that are not attacked.} 

    The investor realizes a \textit{Pareto} improvement by either obtaining a higher level of protection for the same investment\index{investment} or by obtaining the same level of protection at a lower cost; because fewer resources, such as time and human labor, can be largely substituted by algorithms, and automation might be used. As a result, with the implementation of BDA, cyber-security investment\index{investment} might be significantly reduced by disruption: BDA would introduce a discontinuity in the security-breach probability function\index{security-breach probability function (SBPF)}, hence modifying the original GL model\index{Gordon-Loeb (GL) model} assumption of continuity. Accordingly, adapting the security-breach probability function\index{security-breach probability function (SBPF)} ($S^{I}$) from \cite{gordon_economics_2002}, we propose the following security-breach probability function ($S^{I'}_{i}$) in order to capture the advent of a disruptive technology\index{disruptive technologies} by introducing discountinuity through the parameter $d_{i}$:

	        \vspace{0.5cm}
	    
	    \begin{equation} \label{eq:2}
		    S^{I'}_{i}(z_{i}, v_{i})=\frac{v_{i}}{(\alpha_{i}z_{i}+1)^{\beta_{i}+d_{i}}}
	    \end{equation}
	    
	     	\vspace{0.2cm}

	    \begin{quote}
		    wherein for each period $i \in [1,n]$:
			    \begin{itemize}
				    \item[--] $\alpha_{i} \in \mathbb{R}_{> 0}$ and $\beta_{i} \in \mathbb{R}_{\geq 1}$ represent productivity parameters of a given cyber-security\index{cyber-security} technology\index{information technologies (IT)} at period $i$ (i.e., for a given $(z_{i}, v_{i})$, the security-breach probability function $S^{I'}_{i}$ is decreasing in both $\alpha_{i}$ and $\beta_{i}$);\footnote{ \cite{gordon_economics_2002} did not specify $\alpha$ and $\beta$. These productivity parameters could be, for instance, the relative productivity of a given technology when compared to another ($\alpha$) and the joint productivity of the same technology when in interaction with a set of already employed technologies in an organization ($\beta$).}
				    \item[--] $d_{i}$ is a discontinuity parameter at period $i$, and it is represented by a dummy variable. This dummy takes the value $0$ when no disruptive technology is used, and $1$ otherwise.  
			    \end{itemize}
	    \end{quote}
	    
	    	\vspace{0.2cm}
	    	
	\noindent 
	From equation \ref{eq:2}, equation \ref{eq:1} can be rewritten as: 
	
	 \begin{equation} \label{eq:3}
		    \max_{z_{i} \in Z} \Delta^{-1} ENBIS(z_{i}) = \max_{z_{i} \in Z} {\left\{\sum_{i=1}^{n} [v_{i}-\frac{v_{i}}{(\alpha_{i}z_{i}+1)^{\beta_{i}+d_{i}}}]L_{i}-z_{i}\right\}}
	 \end{equation}
	 
	 	    \vspace{0.5cm}

    Although, in order to extend the original GL model\index{Gordon-Loeb (GL) model}, the multi-period setup and the suggested security-breach probability function\index{security-breach probability function (SBPF)} ($S^{I'}_{i}$) constitute the main theoretical contributions, the application of this contribution to a concrete context is necessary in order to exemplify and demonstrate their relevancy, and to empirically test them in a further research.

\section{Application for CIPs}
\label{Isec4}

A cyber-security\index{cyber-security} breach inflicted on a critical infrastructure\index{critical infrastructures (CI)} (CI) generates massive negative externalities, especially due to the increasing interdependency and to technical interconnectedness of different CIs\index{critical infrastructures (CI)} \cite{gordon_externalities_2015,alcaraz_critical_2015}. As a result, cyber-security\index{cyber-security} issues are the main challenge for CIP\index{critical-infrastructure providers (CIP)} \cite{moore_security_2010}. Therefore, the issue of cyber-security investment\index{investment} becomes highly relevant in the context of CIs\index{critical infrastructures (CI)}, and especially so from a social-welfare perspective \cite{gordon_externalities_2015}. 

In this respect, cyber-security investment is a national security priority for the great majority of governments (e.g., \cite{organisation_2012}). For example, on February 12, 2013, the administration of US President Barack Obama implemented Executive Order 13636 \cite{order_2013} named \textit{Improving Critical Infrastructure Cybersecurity}. This executive order stated that ‘\textit{The national and economic security of the United States depends on the reliable functioning of the Nation’s critical infrastructure in the face of such threats. It is the policy of the United States to enhance the security and resilience of the Nation’s critical infrastructure and to maintain a cyber-environment that encourages efficiency, innovation, and economic prosperity while promoting safety, security, business confidentiality, privacy, and civil liberties}' \cite{order_2013}. Following this trend, on May 11, 2017, the administration of US President Donald Trump implemented Executive Order 13800 \cite{order_2017}, named \textit{Strengthening the Cybersecurity of Federal Networks and Critical Infrastructure}. This executive order stated that ‘\textit{Effective immediately, each agency head shall use ‘The Framework for Improving Critical Infrastructure Cybersecurity' developed by the National Institute of Standards and Technology, or any successor document, to manage the agency’s cyber-security risk. Each agency head shall provide a risk management report to the Secretary of Homeland Security and the Director of the Office of Management and Budget within 90 days of the date of this order}' \cite{order_2017}.\\

We believe that our suggested extension of the GL model should be generalizable across any kind of cyber-security\index{cyber-security} concerns, yet we illustrate our approach in the context of CIP\index{critical-infrastructure providers (CIP)} as security-cost reduction related to risk events are of prior importance because potential losses reach unaffordable dimensions \cite{gordon_externalities_2015}. Despite the fact that BDA is considered a promising method for CIP\index{critical-infrastructure providers (CIP)}, its concrete implications have not been discussed much.

More specifically, our extension of the GL model provides a theoretical framework for analyzing the impact of the implementation of any given novel technology on cyber-security investment \textit{over time}. In this respect, by subtracting antidifference operators of $ENBIS$, $\Delta^{-1}ENBIS(z_{i})$, at two different specified time-span of same duration, ($A-B$), a potential decrease in cyber-security investment, $\Delta_{Z}$, can be analyzed -- \textit{ceteris paribus}:

	        \vspace{0.5cm}
 		    
	    \begin{equation} \label{eq:4}
		    \Delta_{Z}={\left\{\Delta^{-1}ENBIS(z_{A})\right\}}-{\left\{\Delta^{-1}ENBIS(z_{B})\right\}}
	    \end{equation}
	    
	        \vspace{0.2cm}
	    
	    \begin{quote}
            where $B \subseteq \mathbb{N}$ is a time-span, wherein each period $i_{B} \in [n,t]$, and in which a given novel technology is implemented; whereas $A \subseteq \mathbb{N}$ is a time-span, wherein each period $i_{A} \in [1,m]$, and in which no novel technology is implemented. Note that ${\sum_{i=n}^{t}i}$ and ${\sum_{i=1}^{m}i}$ must be equal in order to proceed to the comparison. 
	    \end{quote}
	    
 		    \vspace{0.5cm}
\noindent
From equations \ref{eq:1} and \ref{eq:4}, equation \ref{eq:5} can be written as: 
 		    
	    \begin{equation} \label{eq:5}
		    \Delta_{Z}={\left\{\sum_{i=1}^{m} [v_{i}-S^{I}_{i}(z_{i,}v_{i})]L_{i}-z_{i}\right\}}-{\left\{\sum_{i=n}^{t} [v_{i}-S^{I}_{i}(z_{i,}v_{i})]L_{i}-z_{i}\right\}}
	    \end{equation}
	    
 		   \vspace{0.5cm}

\noindent
From the original equation of \cite{gordon_economics_2002}, equation \ref{eq:5} can be rewritten as:

	    \begin{equation} \label{eq:6}
		    \Delta_{Z}={\left\{\sum_{i=1}^{m} [v_{i}-\frac{v_{i}}{(\alpha_{i}z_{i}+1)^{\beta_{i}}}]L_{i}-z_{i}\right\}}-{\left\{\sum_{i=n}^{t} [v_{i}-\frac{v_{i}}{(\alpha_{i}z_{i}+1)^{\beta_{i}}}]L_{i}-z_{i}\right\}}
	    \end{equation}
	    
 		    \vspace{0.5cm}
 		    
By controlling for every parameters to remain equal between time-span $A$ and $B$ -- except for the domains $Z$, wherein $Z_{A} \neq Z_{B}$ --, an organization can determine if any given novel technology generates a greater $ENBIS$. If it is the case, the organization might consider the novel technology implemented in $B$ as disruptive; inversely, if the $ENBIS$ does not substantially changes, the implemented novel technology cannot be considered as disruptive. Note that even though equation \ref{eq:6} can be used for analyzing the differential in cyber-security investment between two time-spans, equation \ref{eq:3} remains necessary in order to determine the optimal level of cyber-security investment ($z_{i}$) through the maximization of the $ENBIS(z_{i})$ function.

The application of our GL model\index{Gordon-Loeb (GL) model} extension to the context of CIP\index{critical-infrastructure providers (CIP)} should provide us with a relevant and seminal basis on which our arguments can be formally modeled and simulated. In the case of human-processed information and defense tactics, these issues would probably make the optimal level of protection difficult to attain or even impossible to finance, as the expected loss would be extreme in the case of cascading failures\index{cascading failures} of CIs\index{critical infrastructures (CI)}; hence the resulting cyber-security\index{cyber-security} investment\index{investment} would also have to reach extreme levels. However, with the effects that the BDA technology\index{information technologies (IT)} (or any similar disruptive technology\index{disruptive technologies}) could have on investment\index{investment} in cyber-security\index{cyber-security}, the investment\index{investment} needs by CIP\index{critical-infrastructure providers (CIP)} could remain at the same level or even decrease, as these novel threats are neutralized by the superior technology\index{information technologies (IT)} that BDA offers.

\section{Discussion}
\label{Isec5}

In this last section, we present our concluding comments, the policy recommendations resulting from concluding comments, we discuss the limitations of this study and suggest paths for further research.

    \subsection{Concluding Comments}
    In this article, we propose an extension of the GL model\index{Gordon-Loeb (GL) model} by adapting its initial theoretical framework in order to capture time-related insights related to the consequences that a disruptive technology can have on cyber-security investment\index{investment}. We propose two important contributions. First, we argue that a single-period model is not adapted to capture dynamic aspects of cyber-security investment\index{investment} such as the advent of a disruptive technology\index{disruptive technologies}. The extension to a multi-period model is indeed necessary. Second, in the context of the introduction of a discrete disruptive cyber-security technology\index{information technologies (IT)}, the security-breach probability function\index{security-breach probability function (SBPF)} of the original GL model\index{Gordon-Loeb (GL) model} could not be considered as continuously differentiable. These two arguments enrich the initial model by giving supplementary insight on cyber-security investment\index{investment}. Through our extended $ENBIS$ function, our proposed revision of the GL model captures both the financial consequences on the optimal level of investment\index{investment} and the security productivity\index{productivity} of the advent of any given novel (and potentially disruptive) technology\index{disruptive technologies}. Although we believe that this reasoning is generalizable across a wide range of cyber-security\index{cyber-security} concerns, we illustrate our approach in the context of CIP\index{critical-infrastructure providers (CIP)}, for which cyber-security\index{cyber-security} breaches inflict unaffordable social costs that urgently need to be reduced.
    
    \subsection{Considerations for Policy Recommendations}
    First and foremost, our extension of the GL model is intended to be used for determining the optimal level of cyber-security investment through the maximization of the $ENBIS(z_{i})$ function. The optimal level of cyber-security investment is obtained when the difference between benefits and costs are maximized -- i.e., where the marginal benefits of cyber-security investment are equivalent to the marginal cost of potential losses due to cyber-security threats.
        
    Also, our extension of the GL model provides a theoretical framework in order to analyze the expected net-benefits of the implementation of any given novel technology over time. By subtracting antidifference operators of $ENBIS$ at two different specified time-spans, and by controlling for every parameters to remain equal between two time-spans (except for investment), a potential decrease in cyber-security investment can be analyzed. Throughout our model, an organization can determine if any given novel technology generates a greater $ENBIS$. If it is the case, the organization might consider the novel technology implemented as disruptive; inversely, if the $ENBIS$ does not substantially changes, the implemented novel technology cannot be considered as disruptive. 
    
    We presented a dynamic analysis for determining the optimal investment\index{investment} level for IS defense, in the context of potentially disruptive technologies\index{disruptive technologies}. By conceptualizing a security-breach probability function\index{security-breach probability function (SBPF)} that includes productivity parameters ($\alpha$ and $\beta$) of a given cyber-security\index{cyber-security} technology\index{information technologies (IT)}, practitioners can calculate the optimal level of investment\index{investment} in IS defense\index{information-systems (IS) defense capability} processes, products and/or services, and also select them according to the highest productivity parameters. Although this work remains theoretical, it gives a systematic method for optimizing IS defense investment\index{investment} in the context of potentially disruptive technologies\index{disruptive technologies}.

    \subsection{Limitations and Paths for Further Research}
    Finally, our research design -- based on theoretical modeling and utility maximization -- has some limitations that future research could help relax.
    
    First, the productivity parameters ($\alpha_{i}$ and $\beta_{i}$) of a given cyber-security\index{cyber-security} technology\index{information technologies (IT)} are theoretically conceptualized, but not empirically measured. Such an effort could add a substantive benefits in terms of validation/refutation of our propositions. However, researchers might find it difficult to gain access to data as they are substantially difficult to measure and/or to find \cite{moore_2019}. A possible solution in order to measure these productivity parameters could be derived from equation \ref{eq:6}, by computing the ratio between $Z_{B}$ and $Z_{A}$. By doing so, a relative measure of productivity between a technology employed in $\Delta^{-1}ENBIS(z_{A})$ and a technology employed in $\Delta^{-1}ENBIS(z_{B})$ could be extracted.
    
    Second, the term \textit{disruptive technologies\index{disruptive technologies}} \cite{christensen_what_2015} has been qualitatively defined, but not quantitatively delimited. Such a research effort is necessary in order to delimit the dummy variable $d_{i}$ (that takes the value $0$ when no disruptive technology\index{information technologies (IT)} is used, and $1$ otherwise). Again, such a quantitative delimitation could be drawn by determining a threshold value in the aforementioned ratio between $Z_{B}$ and $Z_{A}$.
    
    Third, and consequently, this article formally modeled an extension to the GL model, but it did not simulate or empirically test our suggested extension. We decided to leave this operationalization to future research as we wanted to focus on the generalizability of the model. Any simulation or empirical operationalization requires a specification of the above-mentioned productivity parameters $\alpha_{i}$ and $\beta_{i}$ of each technology considered for investment, as well as the specification of the dummy variable $d_{i}$ for classifying which technologies are considered disruptive (and when). Such choices make the model more specific to particular assumptions about technological and productivity contexts. Hence, we suggest that our model should be operationalized by a series of different simulations and empirical tests, rather than by one illustrative simulation run. Given the fact that specifications of such simulations and/or empirical tests would require multiple cases and thus multiple contexts (and thus multiple articles). Nevertheless, we recommend that future research simulates, tests and develops further the model we proposed here. 
    
    Fourth, the security-breach probability function $S^{I'}_{i}$ that we employed in this work can be replaced by many other families of more complex functions. In this research, we selected the first family of security-breach probability function proposed by \cite{gordon_economics_2002}, namely $S^{I}$. However, in reality, adding more families of functions could enrich the analysis (e.g., \cite{lelarge_coordination_2012}).\\
    
    \noindent
    Once the aforementioned points are defined by empirical analyzes and field surveys, further research could propose to simulate a multi-player and multi-period game (e.g., \cite{grossklags_secure_2008}) that models the cyber-security\index{cyber-security} for CIP\index{critical-infrastructure providers (CIP)} in the era of disruptive technologies\index{disruptive technologies}. Such research -- by collecting simulated data and quantitatively analyzing them -- would contribute to complement the theoretical approach presented in this article, hence test the intuition of our theoretical development.

\bibliographystyle{acm}
\bibliography{mybibliography}

\end{document}